\documentclass{webofc}
\usepackage[varg]{txfonts}   
\begin{document}
\title{Cosmology with the SZ spectrum: \\ measuring the Universe's temperature with galaxy clusters}
%
%

\author{\firstname{Gemma} \lastname{Luzzi}\inst{1}\fnsep\thanks{\email{gemma.luzzi@ssdc.asi.it}} \and 
	\firstname{Emanuele} \lastname{D'Angelo}\inst{2} 
	\and
	\firstname{Herve} \lastname{Bourdin}\inst{3}
	\and
	\firstname{Federico} \lastname{De Luca}\inst{3}
	\and
	\firstname{Pasquale} \lastname{Mazzotta}\inst{3}
	\and
	\firstname{Filippo} \lastname{Oppizzi}\inst{3}
	\and 
	\firstname{Gianluca}
	\lastname{Polenta}\inst{1}
}

\institute{Space Science Data Center, Italian Space Agency, via del Politecnico, 00133, Roma, Italy 
\and
           Dipartimento di Matematica, Università degli Studi di Roma “Tor Vergata,” \\ via della Ricerca Scientifica, 1, I-00133 Roma, Italy
\and
           Dipartimento di Fisica, Università degli Studi di Roma “Tor Vergata,” via della Ricerca Scientifica, 1, I-00133 Roma, Italy
          }

\abstract{%
  The hot gas in clusters of galaxies creates a distinctive spectral distortion in the cosmic microwave background (CMB) via the Sunyaev-Zel'dovich (SZ) effect. The spectral signature of the SZ can be used to measure the CMB temperature at cluster redshift ($T_{\rm CMB}(z)$) and to constrain the monopole of the $y$-type spectral distortion of the CMB spectrum. In this work, we start showing the measurements of $T_{\rm CMB}(z)$ for a sample extracted from the Second Catalog of galaxy clusters produced by Planck (PSZ2) and containing $75$ clusters selected from CHEX-MATE. Then we show the forecasts for future CMB experiments about
the constraints on the monopole of the y-type spectral distortion of the CMB spectrum via the spectrum of the SZE.
}
\maketitle
%
\section{Introduction}
\label{intro}
The COBE-FIRAS experiment revealed that the average CMB spectrum is extremely close to a perfect black-body with present-day temperature $T_0=2.7260 \pm 0.0013 ~\rm K$ \cite{fixsen09} and with possible distortions $< 10^{-5}$. Departures of the CMB frequency spectrum from a pure black body encode information about the thermal history of the early Universe \cite{chluba2018future}.
According to the Big Bang model, under the assumptions of adiabatic expansion and photon number conservation, the CMB temperature evolves with redshift $z$ as
$T_{\rm CMB}(z) = T_0(1+z)$
which is one of the core predictions of standard big bang cosmology that may be violated under nonstandard scenarios \cite{Tofz}.
Testing its validity is important for cosmology and fundamental physics. The possibility of determining $T_{\rm CMB}(z)$, $T_0$ and the monopole of $y$-type distortion of the CMB spectrum from measurements of the Sunyaev–Zeldovich effect had been suggested long ago \cite{fabbri78,rephaeli80}.
Building upon these two works,
in Sect.~\ref{sec-1}) we present constraints on the CMB temperature evolution with redshift and independent estimate of $T_0$. In Sect ~\ref{sec-2} we show forecasts for future CMB experiments about constraints on the $y$-type distortion of the CMB spectrum.

\section{Constraining the evolution of the CMB temperature for the CHEX-MATE sample of Planck SZ clusters}
\label{sec-1}
Deviations of the standard CMB temperature scaling with redshift are usually described using the parametrization proposed by \cite{lima00},
$T_{\rm CMB}(z) = T_0(1+z)^{(1-\beta)}$, 
where $\beta$ is a constant parameter ($\beta = 0$ in the standard scenario).
There have been various attempts to constrain the CMB temperature evolution via the SZ method. Starting from the initial constraints using a few clusters of \cite{battistelli02,luzzi09}, notable improvements in constraining the $\beta$ value are obtained by using large galaxy cluster catalogues together with very precise CMB data from the Planck satellite \cite{demartino15,hurier14,luzzi15}, from the South Pole Telescope \cite{saro14}
and from The Atacama Cosmology Telescope \cite{Li:2021hxe}. The core of the present analysis rests upon a sample of the second Planck SZ cluster catalog (PSZ2) \cite{PSZ2_2015} selected by CHEX-MATE \cite{heritage2021_a} and Planck PR2 temperature maps\footnote{\url{http://www.sciops.esa.int/index.php?project=planck\&page=Planck\_Legacy\_Archive}} at frequencies from 100 to 857 GHz \cite{planck2015results_I}. For this sample we derive information about the $\rm R_{500}$ and about the electronic temperature $T_e$, which will be included as prior information in the subsequent analysis for the extraction of the CMB temperature at the redshift of each cluster.
Here we show results only for 77 clusters in the CHEX-MATE sample. The analysis for full sample, consisting of $118$ clusters, will be presented in D'Angelo et al in preparation.

\subsection{SZ maps at Planck frequencies and modelling of the SZ signal} \label{subsec-szmaps}
 It is widely known that the \textit{Planck} maps include radiation from a variety of astrophysical sources, chiefly the CMB, the SZ signal, Galactic foregrounds, CO emission, zodiacal light and point sources. In order to separate these components from the tSZ signal we apply to the \textit{Planck} PR2 data the same recipe used by \cite{bourdin2017} on PR1 data. In particular CMB and dust anisotropies are spectrally modelled using all HFI frequency maps outside each galaxy cluster ($7\rm R_{500}$ $\leq$ $\rm R$ $\leq$ $12\rm R_{500}$), and spatially modelled using a linear combination of wavelet filtered frequency maps at 857 and 217 GHz. SZ signal maps at 100, 143, 217 and 353GHz are used to build the likelihood for the $T_{\rm CMB}(z)$ extraction. For each cluster, we assume that the pressure structure follows a spherically symmetric distribution, $P(r)$, first proposed by \cite{nagai2007}. Apart from the amplitude, which is left free to vary, and the size, which is determined by the analysis of the CHEX-MATE data, all the other parameters of the pressure profile are fixed at the values of \cite{arnaud2010}. We can thus factorize the SZ signal of a cluster into two terms: the SZ template and the SZ amplitude $\Delta T_{\mathrm{SZ}}(\nu,\Theta_n)$ which depends on the unknown parameters we want to extract. Assuming that the SZ data are gaussianly distributed allows us to construct the single cluster likelihood function. The likelihood for the $n$-th cluster is:

\begin{equation}\label{eq:likelihood_deltaCMB}
	P(\mathrm{map}_{\mathrm{obs}}| \Theta_n)
	  \propto  \exp \left\{ -\sum_{\nu}\sum_{ij} \frac{[\Delta T_{\mathrm{SZ}}(\nu,\Theta_n)\times\mathrm{template }(\nu,i,j)-\mathrm{map}_{\mathrm{obs}}(\nu,i,j)]^2}{\mathrm{varmap}(\nu,i,j)+\delta {CMB}^2} \right\}
\end{equation}
where $\Theta_n$ are the cluster parameters (Comptonization parameter $y_n$, peculiar velocity $v_{pec}$, electron temperature $T_e$) and the CMB temperature at the redshift of the cluster, while $\mathrm{varmap}(\nu,i,j)$ is the variance in thermodynamic temperature for each pixel and for each frequency channel, as provided by the Planck collaboration.  
 Kinematic SZ and the CMB primary anisotropies have the same spectral shape in the non-relativistic limit, thus in cleaning for CMB we also remove the main kinematic component of the SZ. We account for possible CMB and KSZ residuals in our maps after the cleaning procedure. We model the CMB residual signal as in \cite{DAgostini:2000uvy}, accounting for a systematic effect which is a not exactly known offset, the uncertainty about which is described by a Gaussian pdf around zero and standard deviation $\delta \rm CMB$ (estimated on a set of simulations). While to account for possible KSZ residuals, we model them as a kinematic SZ component and adopt as a prior a Gaussian with a universal vanishing mean and with a 500 km/s standard deviation.

\subsection{$T_{\rm CMB}(z)$ with the MCMC sampler Cobaya}
In order to extract $T_{\rm CMB}(z)$ we adopt the publicly available MCMC sampler Cobaya\footnote{https://cobaya.readthedocs.io/en/latest/} \cite{Torrado_2021,torrado2019}, able to reconstruct the posterior distribution for the parameters of interest. The MCMC chains were analyzed by using GetDist\footnote{\url{https://github.com/cmbant/getdist}}. We explore the space of the cluster parameters (Comptonization parameter, electron temperature Te) and the CMB temperature at the redshift of the cluster. 
\begin{figure}\label{figcontourfixte}
\centering
\includegraphics[scale=0.3]{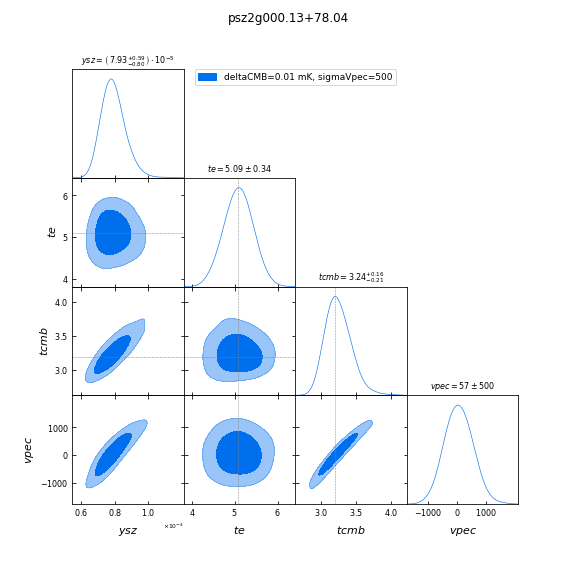}
\caption{1D and 2D posteriors for one cluster of the sample. We adopt a gaussian prior for the electronic temperature, while keeping free the CMB temperature parameter and the comptonization parameter.  Vertical grey lines denote respectively: for $T_e$ the central value from X-ray data, for $T_{CMB}$ the value expected by the standard scaling. }
\end{figure}
Examples of parameter extraction performed on SZ maps are provided in figure 1
, where the gaussian prior on the electronic temperature is derived by the analysis of the CHEX-MATE data.

\subsection{CMB temperature measurements and constraints on the $\beta$ parameter}
\begin{figure}
\centering
\includegraphics[scale=0.25]{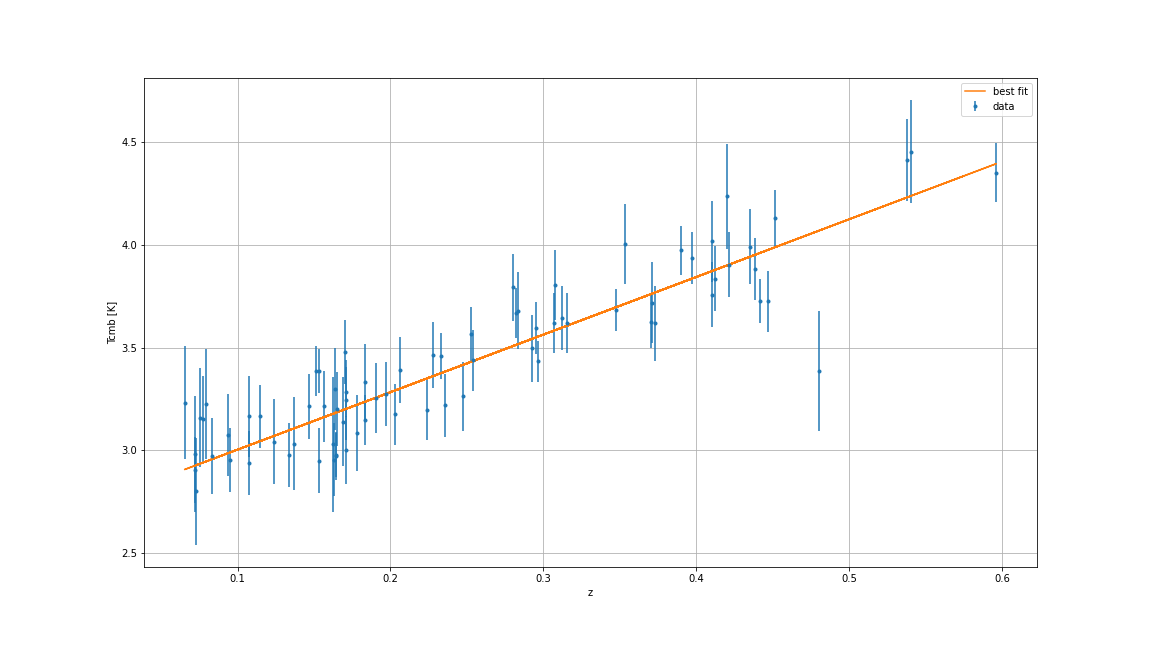}
\caption{Measurements of the CMB temperature as a function of redshift for 77 \textit{Planck} selected clusters. The orange line is the best fit (with $\beta = -0.02$) to the scaling $T_{\rm CMB}(z)$=$T_0(1+z)^{1-\beta}$.}
\label{fig:TcmbReal}       
\end{figure}

In figure~\ref{fig:TcmbReal} we plot the measurements of the CMB temperature as a function of redshift. We report expected values and standard deviation of each single $T_{\rm CMB}(z)$ distribution. Our measurements reach a precision of up to $3\%$.
By fitting the $T_{\rm CMB}(z)$ data with the parametrization $T_{CMB}(z)$=$T_0(1+z)^{1-\beta}$ we get for 77 clusters $\beta = -0.02 \pm 0.02$, at 1$\sigma$ uncertainty, in agreement with the predictions of the standard model. 

\subsection{Independent estimate of $T_0$}
Lately there has been a renewed interest in independent estimate of $T_0$. Motivated by the works of \cite{PhysRevD.102.063515,PhysRevD.103.L081304,PhysRevD.104.043516,2020,refId0},
assuming the standard evolution of the CMB temperature with redshift, we can use our measurements of $T_{\rm CMB}(z)$ to put constraints on the value of $T_{\rm CMB}$ at redshift zero.
By considering $T_0$ as a free parameter when fitting the data with the standard relation $T_{\rm CMB}(z) = T_0(1+z)$, we get an independent estimate of the CMB temperature at redshift zero: $T_0 = 2.729\pm 0.014~ \rm K$, at $1\sigma$ uncertainty, a value consistent with the COBE-FIRAS measurement, $T_0 = 2.7260 \pm 0.0013 ~ \rm K$.

\section{Constraining the monopole $y$ distortion via the SZ spectrum}\label{sec-2}
\begin{figure*}
\centering
\includegraphics[scale=0.6]{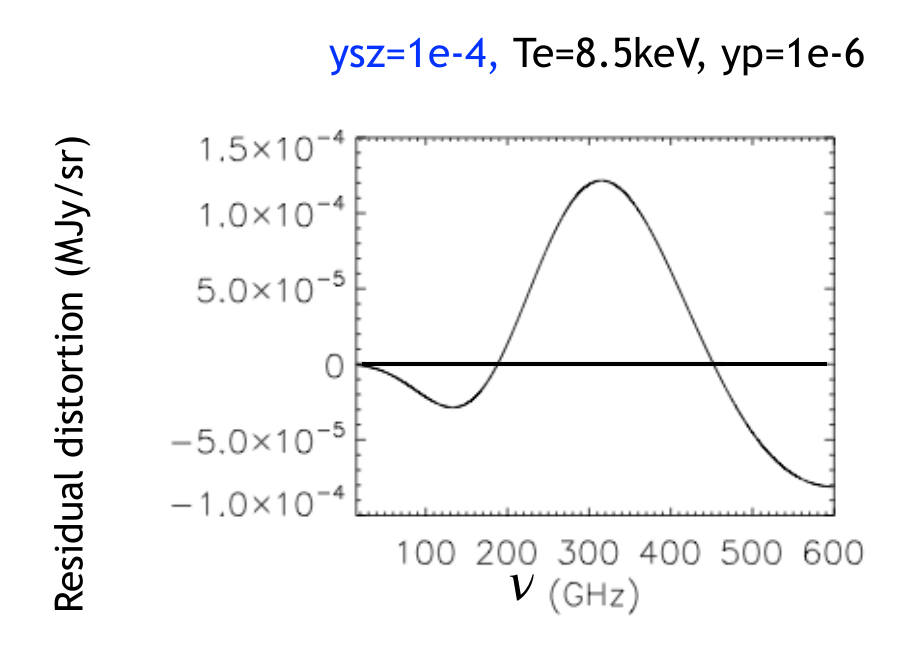}
\includegraphics[scale=0.18]{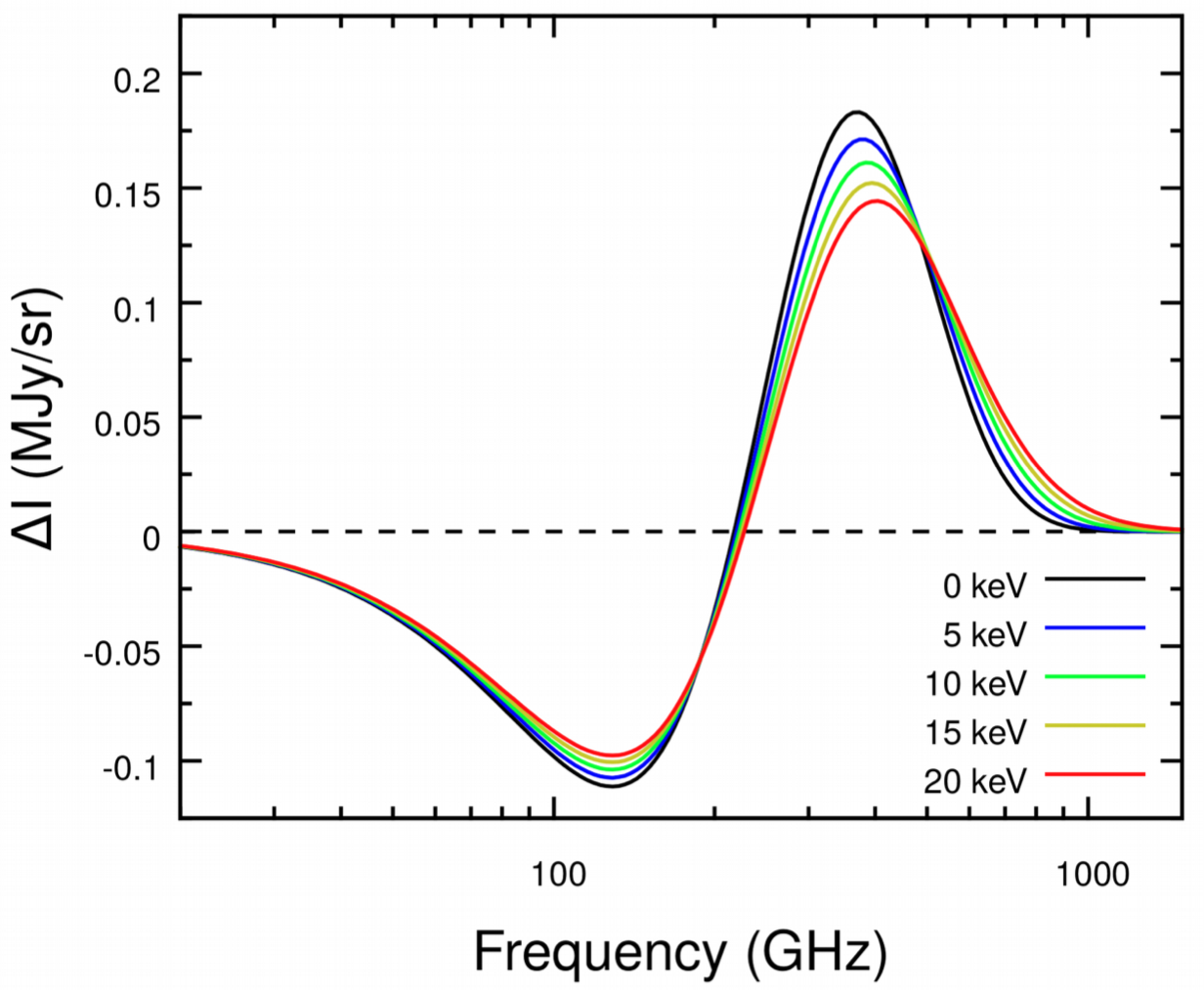}
\caption{On the left: Departures from an SZ spectrum (for a cluster with $y_{SZ}= 10^{-4}$, $T_e=8.5$keV) due to a CMB spectrum with distortion $y_p = 10^{-6}$. On the right: SZ spectra (tSZ with relativistic corrections) for typical $T_e$ values.}
\label{fig:distortion}      
\end{figure*}

Departures of the CMB frequency spectrum from a pure black body encode information about the thermal history of the early Universe.
Notably, spectral distortions of the CMB can be created, after thermalization becomes inefficient at redshifts $z\lesssim 10^6$, by energy-releasing processes that can drive matter and radiation out of thermal equilibrium. Examples include injection of photons or other particles that interact electromagnetically. The effects associated with such processes may be described as y-type and $\mu$-type distortions, characterising the exchange of energy between photons and electrons through Compton scattering. In particular, as Compton scattering becomes inefficient for $z\lesssim 10^4$,
$y$-type distortions may be induced in these redshifts, and so the detection of a primordial $y$-distortion signal, $y_p$, would probe the thermal history of our Universe during the recombination and reionization eras. 
Here we apply the method proposed by \cite{fabbri78,rephaeli80} to constrain the CMB $y_p$ distortion via the SZ spectrum. 
The expected residual distortion due to a CMB $y_p = 10^{-6}$ for a cluster with $y_{SZ}= 10^{-4}$, $T_e=8.5$keV is shown in figure~\ref{fig:distortion}.
Given the amplitude of the signal it is clear that only future CMB experiments could be able to improve the constraints on the CMB $y_p$ distortion via the SZ spectrum. 

\subsection{Forecasts for Millimetron}
\label{subsec-2-1}
\begin{figure}
\centering
\includegraphics[scale=0.27]{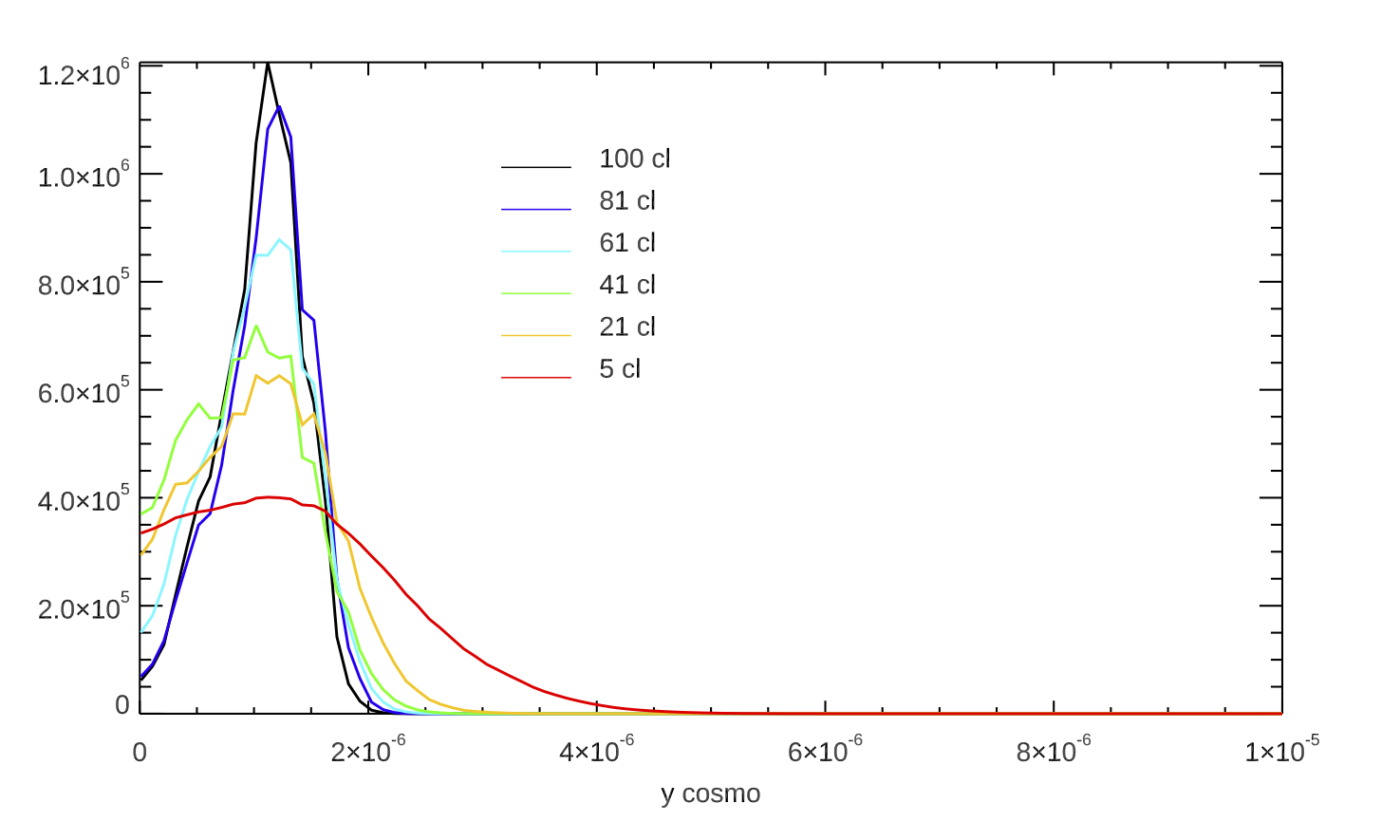}
\caption{Joint pdf on the parameter $y_p$ from simulated Millimetron SZ spectra (5 to 100 clusters considered).}
\label{fig:posterior_yp}      
\end{figure}

The Millimetron Space Observatory (MSO)
\footnote{\url{http://millimetron.ru/en/}}
 is a large, cryogenic space observatory covering the spectral range from 70$\mu m$ to 10mm. The antenna has an opening of 10m and is cooled below 10K. The observatory will work from deep space, both as a single telescope and as part of a very long baseline interferometer (VLBI). In 2011, ASI funded a phase A study for the Long wave-Array Camera Spectrometer (LACS) (PI P. de Bernardis), a focal plane instrument comprising a Fourier transform spectrometer with non-coherent detector mosaics. The main characteristics of the telescope-LACS combination are the following: spectral coverage: 100 GHz - 1000 GHz (in 4 sub-bands); maximum spectral resolution: 1 GHz; angular resolution: from 10 arc seconds to 1 arc minute; field of view about 5 arc minutes; sensitivity: about 100 Jy / sr in a 10 GHz band in 1 hour of integration. With LACS, both differential and absolute spectral measurements are possible. The SZ effect in galaxy clusters constitutes the main scientific goal for LACS. By exploiting the very accurate spectroscopic capabilities of the instrument, as well as its high angular resolution, it will be possible to perform a reliable separation of all spectral components. In figure~\ref{fig:posterior_yp} we show forecasts for the extraction of the CMB $y_p$ distortion via the SZ spectrum for a number of Millimetron clusters ranging from 5 to 100. Even if the simulation adopted are quite simplistic, since the SZ spectra in the range 100 - 600 GHz are recovered assuming perfect cleaning of foregrounds, it seems that this differential approach can be complementary to absolute spectroscopic measurements of the CMB y distortions.

\bibliography{template}

\end{document}